\documentclass[twoside]{article}
\usepackage{fleqn,espcrc2,epsfig}

\def \beq {\begin{equation}}
\def \eeq {\end{equation}}
\def \be {\begin{eqnarray}}
\def \ee {\end{eqnarray}}
\def \ben {\begin{eqnarray*}}
\def \een {\end{eqnarray*}}

\newcommand{\AmS}{{\protect\the\textfont2
  A\kern-.1667em\lower.5ex\hbox{M}\kern-.125emS}}

\title{Vortices in $SU(2)$ lattice gauge theory}

\author{S.~Cheluvaraja\thanks{Work supported by the
U.S. Department of Energy under grant
DE-FG05-91 ER 40617.}\address{Dept. of Physics and Astronomy,
        Louisiana State University, Baton Rouge, Louisiana,
        70803  USA}}

\begin{document}

\begin{abstract}
We investigate some properties of thick vortices and thick monopoles
in the $SU(2)$ lattice gauge theory
by inserting operators that
create these excitations.
We measure the
derivative of the free energy
of the vortex
with respect to the coupling
and we find that it falls exponentially with the cross sectional
area of the vortex.
We also study the monopole-monopole potential energy of thick
and thin monopoles.
Our results suggest that vortices and monopoles of increasing thickness
will play an important role in the large $\beta$ limit.
\vspace{1pc}
\end{abstract}

\maketitle

Vortices as a mechanism for quark confinement have attracted enormous interest
in recent times. In non-abelian gauge theories, center vortices can be
defined as
multiple valued gauge transformations. Furthermore, it has been conjectured
that center vortices can condense  and produce quark confinement \cite{hooft1}.
Vortices can be introduced by applying twisted boundary
conditions (in the spatial direction) \cite{hooft2}. 
On the lattice, the twist translates to flipping
the signs of a stack of plaquettes in a particular spatial direction, the
same twist repeating at every time slice. Applying this twist corresponds to
inserting a center vortex into the system. Recent work in \cite{tombo}
studied the free energy of a vortex, whereas the authors in \cite{dgc} have
constructed a vortex disorder operator and measured its expectation value.
Vortices have also been studied in the center projection approach \cite{proj1}.

When one tries to define vortex like configurations on the lattice one is
led to a distinction between thin and thick vortices \cite{mack}.
Thick vortices are configurations having $sign(trU(p))=+1$ everywhere
but which nevertheless flip signs of Wilson loops linking them. Thin
vortices, on the other hand, are configurations of co-closed sets of
plaquettes with $sign(trU(p))=-1$ and are defined at the scale of the lattice spacing. The two configurations are not only quite different, but also have
different  properties. It is a dynamical issue whether one is more important
than the other. In this discussion we study some properties of thick
vortices and thick monopoles by generalizing the idea of a twist.

We introduce a term in the partition function that creates a stress
in the system and we study the effect of this stress on the system.
More specifically, we add a term to the Wilson action
\beq
S^{\prime}=\frac{\beta}{2}\sum_{p}trU(p) -\frac{\beta}{2}\sum_{p^{\prime}}trU(d)
\quad .
\eeq
The extra term is a Wilson loop variable defined over a square of side $d$ and 
it comes with
a negative sign. This term is introduced at a point in the (say) $ (12)$
plane, and it extends in the $3$ direction, and the term repeats
itself in the $4$ direction.
The effect of the extra term is to drive the system such that
$trU(d)$ is negative, the effect of the usual Wilson action being to drive
the system such that $trU(p)$ is positive. 
The stress that
we are introducing into the system is nothing but a thick vortex of thickness
$d$ piercing a region in the $(12)$ plane and wrapping around the lattice
in $3$ direction. On the other hand, if the stress begins at a point, say
$z=z1$, and terminates at, say $z=z2$, we are inserting a thick monopole-antimonopole pair joined
together by a thick vortex line. 
We want to measure the system free energy cost for introducing these
stresses.
The case $d=1$ has been studied before \cite{tombo,oths}, but we will be interested in the free energy cost
as a function of $d$.
Since direct free energy measurements are quite
difficult we appeal to a simpler method to determine the effect of
this stress ( the same idea is used in \cite{dgc}).
We first write
\beq
\mu_{d}=\frac{Z(\beta^{\prime})}{Z(\beta)}=\exp -(F_{d}-F_{o})
\quad .
\eeq
$Z(\beta^{\prime})$ is the partition function of the system after applying the
stress and $Z(\beta)$  is the partition function without the stress.
$F_{d}-F_{0}$ measures the free energy difference when an
\it {additional} \rm thick vortex (having a thickness $d$)
is introduced into the original system (which may already contain thick vortices). Hence the effect of a large number of interacting vortices is already
subsumed
in this free energy difference.
It is not difficult to see that
$\frac{\partial \log \mu_{d}}{\partial \beta}$ is
\be
\frac{1}{\mu_{d}}
\frac{\partial \mu_{d}}{\partial \beta} =<\frac{1}{2} \sum_{p}trU(p) -
\frac{1}{2} \sum_{p^{\prime}} trU(d)>_{1}  \\ \nonumber
-<\frac{1}{2}\sum_{p} trU(p)>_{0}
\quad .
\label{maineq}
\ee
The subscript $1$ indicates that the average is taken with respect to the
stressed partition function, whereas the subscript $0$ indicates that the
average is taken with respect to the unstressed partition function. This
quantity directly measures the derivative of the free energy difference
because
\beq
\frac{\partial \log \mu_{d}}{\partial \beta}= \frac{-\partial(F_{d}-F_{0})}{
\partial \beta}
\quad .
\eeq

\begin{figure}[h]
\vspace{9pt}
\framebox[55mm]{\rule[-21mm]{0mm}{43mm}}
\epsfig{file=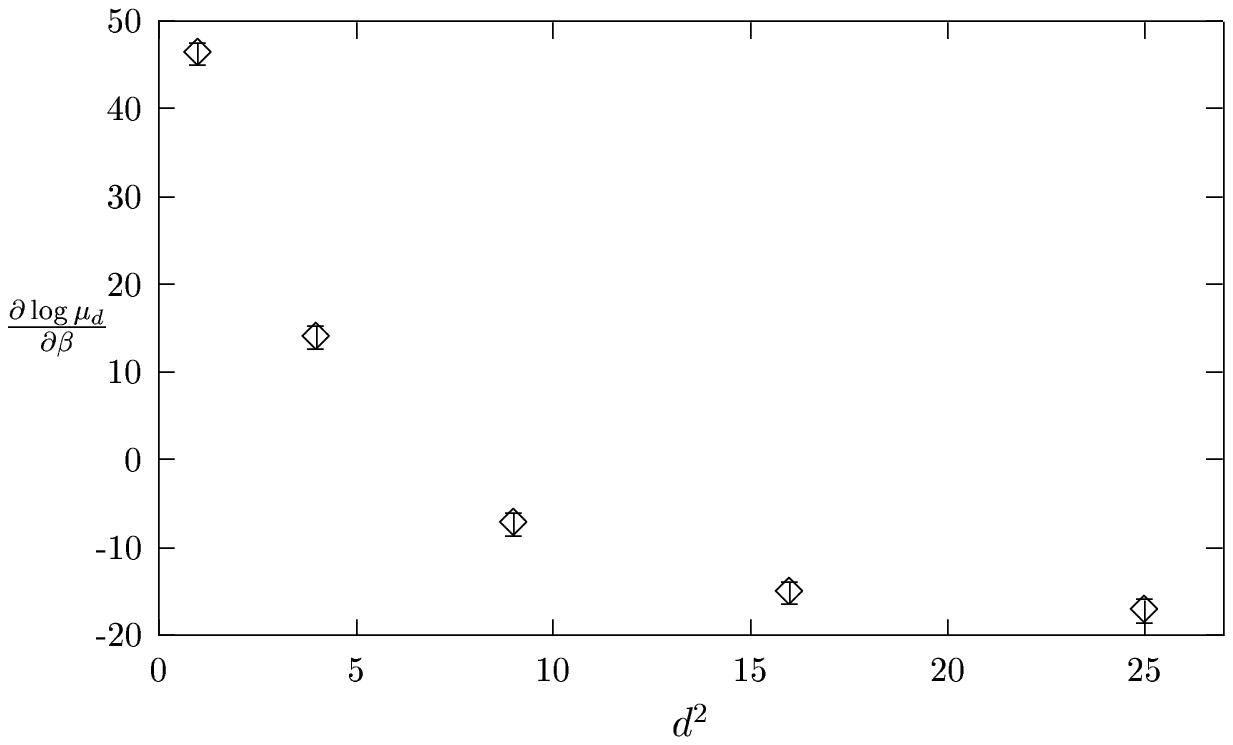,width=7.5cm}
\caption{ $\frac{\partial log \mu}{\partial \beta}$ vs $A$ at $\beta=2.3$.}
\label{2.3}
\end{figure}

Fig.~\ref{2.3} shows
$\frac{\partial log \mu_{d}}{\partial \beta}$ as a function
of the area of the thickness of the vortex. There is an
exponential fall-off with respect to the area of the vortex. A logarithm of
the curve plotted vs the area of the vortex gives an approximately good
straight line fit. The simulations were done on a lattice whose
spatial extent was $10X10$ and the remaining size was $6X6$. Vortices
of thickness ranging from $1$ to $5$ were studied, and in
order to avoid finite size effects associated with these vortices
the lattices were chosen to have a spatial extent of atleast twice the
largest vortex size. The other parameters of the lattice were fixed by
some limitations in the computer time. 
300,000 data points
were gathered at each point and the errors were estimated by binning.

We wish to point out that though our result is for
the derivative of the free energy of the vortex area as a function of the
coupling, the free energy of the vortex will also fall-off exponentially with
the area ( this is very easy to show in the strong coupling limit).
This result has two immediate implications. 
It indicates that as we approach
the weak coupling limit ($\beta \rightarrow \infty$), vortices of 
larger and larger area are more favourable than vortices of any fixed
area. By spreading the core of the vortex over a large area, the
energy of the vortex loops can be made arbitrarily (exponentially) small. 
It also means that in the region of the continuum limit we must
be able to tackle a many body vortex problem in
which vortices can have large overlaps with each other. This feature seems to
make the study of such vortex gases quite difficult. However,
from the point of view of the continuum limit, very large vortices are
almost necessary;
if lattice vortices are
to survive as continuum excitations they must appear with a length scale
which diverges in lattice units  as the lattice spacing goes to zero. Only then can they
yield a vortex corresponding to 
some physical thickness. Thick vortices indeed admit
this possibility by appearing with arbitrarily large thicknesses.

We now turn to thick monopoles. As mentioned earlier, if the stress is
taken to extend only between two points, say $z=z1$ and $z=z2$, it corresponds
to an external thick monopole-monopole loop running along the $4$ direction
at $z=z1$ and $z=z2$ and separated by a thick vortex sheet between $z=z1$ and
$z=z2$. 

\begin{figure}[h]
\framebox[55mm]{\rule[-21mm]{0mm}{43mm}}
\epsfig{file=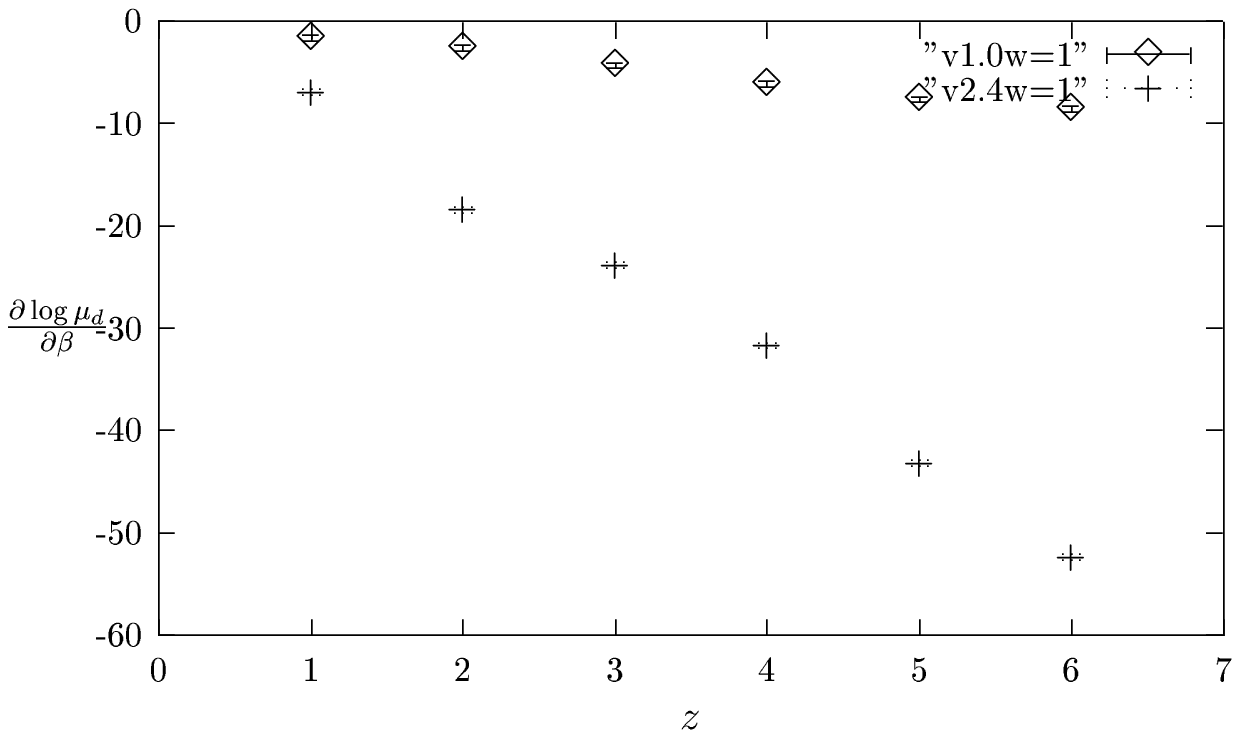,width=7.5cm}
\caption{ Potential energy of thin $d=1$ monopoles as a function of
distance at $\beta= 1.0$ and $\beta=2.3$.}
\label{monpotw=1}
\end{figure}

\begin{figure}[h]
\framebox[55mm]{\rule[-21mm]{0mm}{43mm}}
\epsfig{file=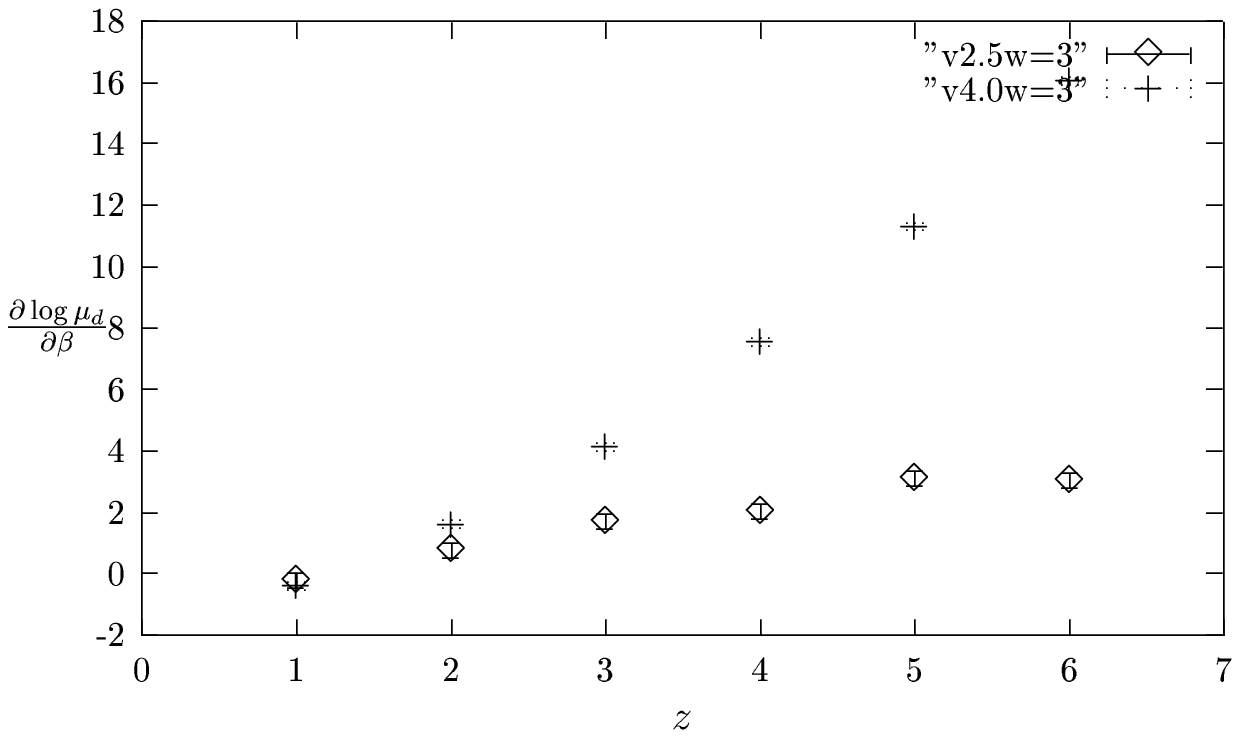,width=7.5cm}
\caption{ Potential energy of thick $d=3$ monopoles as a function of
distance at $\beta=2.4$ and $\beta=4.0$.}
\label{monpotw=3}
\end{figure}

Fig.~\ref{monpotw=1} shows the potential energy of a
monopole-monopole pair for monopoles  
of thickness $d=1$ at $\beta =1.0$ and $\beta=2.4$.
The approximately linear rise at $\beta=2.4$ shows that
their separation energy directly increases with the separation distance.
By contrast, the curve for $\beta=1.0$ is quite flat.
At $\beta=2.4$, the thin monopoles have become very rare and the screening
mechanism between an external monopole-monopole pair is inoperative and their
interaction energy is directly proportional to the length of vortex
line joining them. On the other hand, at $\beta=1.0$, when the monopole density is quite
large, the energy of the external pair does not increase linearly with
their separation because of the screening effect of the 
large number of monopoles in the system.
A similar measurement for
thick monopoles shows very different behaviour.
Fig.~\ref{monpotw=3}
shows the monopole-monopole potential energy for a thick monopole
of thickness $d=3$.
At $\beta=2.4$ the thick
monopole-monopole potential energy curve is quite flat. This can again
be understood from the presence of a large number of thick monopoles ($d=3$)
in the system. 
At larger values of
$\beta$, for example $\beta=4.0$, the thick monopoles also show a
screening behaviour. The main difference between Fig.~\ref{monpotw=1}
and Fig.~\ref{monpotw=3} is the signs of the slopes of the curves in the
two plots. The slope of the curve is the negative of the 
derivative of the free energy of the monopole pair with respect to the
coupling.
While the first plot (with negative slope) shows a linear rise 
with the separation distance in the interaction
energy of a thin monopole-monopole pair at $\beta=2.4$, the second plot ($\beta=4.0$) 
shows a decrease (with positive slope) in the interaction. 
This different behaviour of monopoles of finite
thickness compared with monopoles of unit thickness again shows that the self-
energy of thick monopoles has a more subtle behaviour than that of the
thin monopoles. An appealing explanation of this difference is that
thick monopoles tend to repel each other (like similarly charged particles)
and lower their potential energy. This should be contrasted with the
behaviour of thin monopoles which tend to be bound into monopole-monopole
pairs as a result of the energy of the string joining them.
The data in Fig.~\ref{monpotw=1} and
Fig.~\ref{monpotw=3} were taken on a $6^{4}$ lattice after 50000 iterations.

It is also interesting, and important, to see how
these thick objects scale as we decrease the lattice spacing. We expect that
as the lattice spacing is reduced, objects that were thick on the larger
lattice become thin on the smaller lattice. The time constant of the
exponential in Figs.~\ref{2.3}  should also scale in the large $\beta$ limit.
These issues are being currently studied.

\noindent
Acknowledgement: This work was supported in part by United States
Department of Energy grant DE-FG 05-91 ER 40617.


\end{document}